\documentclass[twocolumn,prb,superscriptaddress,floatfix,aps,10pt]{revtex4-2}
\pdfoutput=1

\usepackage[utf8]{inputenc}
\usepackage[caption=false,subrefformat=parens,labelformat=parens]{subfig}
\usepackage{graphicx}
\usepackage{float}
\usepackage{amsmath}
\usepackage{amssymb}
\usepackage{dcolumn}
\usepackage{bm}
\usepackage{xcolor}
\usepackage{xfrac}
\usepackage[colorlinks=true, allcolors=blue]{hyperref}
\usepackage{chemformula}
\usepackage[normalem]{ulem}
\usepackage{siunitx}
\usepackage{multirow}

\DeclareSIUnit\angstrom{\text {Å}}

\newcommand{\Tc}{T_{\mathrm{c}}}
\newcommand{\Rwp}{R_{\mathrm{wp}}}
\definecolor{mag}{RGB}{255,0,255}

\begin{document}

\title{{Effect of low-temperature compression on superconductivity\\and crystal structure in strontium metal}}
\author{J. Lim}
\email{Corresponding author: jlim5@eiu.edu}
\affiliation{Department of Physics, University of Florida, Gainesville, Florida 32611, USA}
\author{S. Sinha}
\affiliation{Department of Physics, University of Florida, Gainesville, Florida 32611, USA}
\author{D. E. Jackson}
\affiliation{Department of Physics, University of Florida, Gainesville, Florida 32611, USA}
\author{R.\ S.\ Kumar}
\affiliation{Department of Physics, University of Illinois Chicago, Chicago, Illinois 60607, USA}
\author{C.\ Park}
\affiliation{HPCAT, X-ray Science Division, Argonne National Laboratory, Argonne, Illinois 60439, USA}
\author{R.~J.~Hemley}
\affiliation{Departments of Physics, Chemistry, and Earth and Environmental Sciences, University of Illinois Chicago, Chicago, Illinois 60607, USA}
\author{D. VanGennep}
\affiliation{Department of Physics, University of Florida, Gainesville, Florida 32611, USA}
\author{Y.\ K.\ Vohra}
\affiliation{Department of Physics, University of Alabama at Birmingham, Birmingham, Alabama 35294, USA}
\author{R. G. Hennig}
\affiliation{Department of Materials Science and  Engineering, University of Florida, Gainesville, Florida 32611, USA}
\affiliation{Quantum Theory Project, University of Florida, Gainesville, Florida 32611, USA}
\author{P. J. Hirschfeld}
\affiliation{Department of Physics, University of Florida, Gainesville, Florida 32611, USA}
\author{G. R. Stewart}
\affiliation{Department of Physics, University of Florida, Gainesville, Florida 32611, USA}
\author{J. J. Hamlin}
\affiliation{Department of Physics, University of Florida, Gainesville, Florida 32611, USA}
\date{\today}

\begin{abstract}
The superconducting and structural properties of elemental strontium metal were investigated under pressures up to \SI{60}{GPa} while maintaining cryogenic conditions during pressure application.
Applying pressure at low temperatures reveals differences in superconducting and structural phases compared to previous reports obtained at room temperatures.
Notably, the superconducting critical temperature exhibits a twofold increase under compression after cryogenic cooling within the pressure range of 35-\SI{42}{GPa}, compared to cryogenic cooling after room-temperature compression.
Subsequently, the transition width becomes significantly sharper above \SI{42}{GPa}.
Low-temperature X-ray diffraction measurements under pressure reveal that this change corresponds to the Sr-III to Sr-IV transition, with no evidence of any metastable structure.
Furthermore, the monoclinic Sr-IV structure was observed to remain stable to much higher pressures - at least up to \SI{60}{GPa}, without the appearance of the incommensurate Sr-V phase present at room temperature.
This implies that thermal activation energy plays an important role in overcoming the presence of a kinetic barrier to the Sr-V phase at room temperature.
\end{abstract}

\maketitle

\section{Introduction} 
Pressure is a commonly used thermodynamic parameter to influence the structural properties of a material~\cite{Tse_HPCristalReview_2020}.
However, the structural modification can vary depending on the temperature at which pressure is applied~\cite{Cannon_ElementReview_1974}.
This variation arises due to changes in the thermodynamic pathway leading to phase stability, which is closely related to both temperature and pressure.
Applying pressure at low temperatures can lead to the formation of new metastable phases that are distinctly different from the structural trends observed at room temperatures~\cite{Ackland_Lifcc_2017}.
Studies involving low-temperature pressurization below \SI{140}{K} have identified a new barium phase, Ba-VI, which exhibits an orthorhombic structure ($Pnma$) within a pressure range of 13 to \SI{35}{GPa}~\cite{Desgreniers_BaStructure_2015}.
The superconductivity in this Ba-VI phase was extensively investigated by Jackson $et$ $al.$~\cite{Jackson_BaSC_2017}, revealing a maximum superconducting critical temperature ($\Tc$) near \SI{8}{K}, which is twice that observed under high-pressure cooling, $i.e.$ cryogenic cooling after room-temperature compression.

Among the alkaline earth elements, strontium (\ch{Sr}) shares many similarities with barium (\ch{Ba}), particularly in their complex structures.
Notably, both elements exhibit intricate structures such as the incommensurate Sr-V~\cite{McMahon_SrV_2000} and Ba-IV phases~\cite{Nelmes_BaIV_1999, Loa_BIVc_2012}, comprised of two interpenetrating components (host and guest) that exhibit incommensurability along the $c$-axis at room temperature.
Previous structural studies under room-temperature compressions have identified several phases of strontium: Sr-I (fcc, $Fm\bar{3}m$, 0-\SI{3.5}{GPa})~\cite{McWhan_bccSr_1963}, Sr-II (bcc, $Im\bar{3}m$, 3.5-\SI{26}{GPa})~\cite{McWhan_bccSr_1963}, Sr-III ($\beta$-tin, $I4_1/amd$, 26-\SI{35}{GPa})~\cite{Olijnyk_SrIII_1984, Allan_SrIII_1998}, Sr-IV (monoclinic, $Ia$, 35-\SI{46}{GPa})~\cite{Bovornratanarak_SrIV_2006}, Sr-V (incommensurate, host-$I4/mcm$, guest-fct (face centered tetragonal), 46-\SI{74}{GPa})~\cite{McMahon_SrV_2000}.
Of particular interest is the Sr-IV phase, which exhibits a unique monoclinic structure characterized by a distortion of the tetragonal $\beta$-tin Sr-III structure.
This distortion results in a superstructure three times the size of the Sr-III unit cell, featuring a helical chain along the previous Sr-III $c$-axis.
This phenomenon is the result of rather small translations of atoms~\cite{Katzke_BaSrHostGuest_2007, Jackson_Thesis_2018}.

Previous studies on the superconductivity of \ch{Sr} have primarily been conducted under room-temperature compression followed by nearly isobaric cooling~\cite{Dunn_SrBaSC_1982, Mizobata_SrSC_2007, Bireckoven_SrBaSC_1989, Hamlin_ElementsHP_2015}.
Similar to \ch{Ba}, the emergence of pressure-induced superconductivity in \ch{Sr} is attributed to a significant $s$ $\rightarrow$ $d$ electron charge transfer with increasing pressure, supported by band structure calculations~\cite{Skriver_Srndcal_1982, Skriver_ndcalculations_1985,Hamlin_ElementsHP_2015}.
Interestingly, \ch{Sr} exhibits a step-like increase in the $\Tc(P)$ curve during the Sr-IV to Sr-V transition above \SI{46}{GPa} during cryogenic cooling after room-temperature compression.
Considering the discovery of the new metastable phase Ba-IV under low-temperature compression, which demonstrates a distinct $\Tc$ compared to that observed under high-pressure cooling, it is plausible that the same effect of low-temperature compression could lead to the formation of different crystal structures with unique superconducting properties in \ch{Sr}.
Despite extensive prior research, investigations into the superconducting and structural properties of \ch{Sr} have largely focused on high-pressure applications at room temperature.
In this study, we explore low-temperature compression paths to complement the past studies and gain deeper insights into the structural phases and superconductivity of elemental \ch{Sr} metal.

\section{Methods}
For high-pressure electrical resistivity measurements, a 5N high-purity polycrystalline \ch{Sr} sample was loaded into a membrane-driven diamond anvil cell (OmniDAC from Almax-easyLab) and placed inside a customized continuous-flow cryostat (Oxford Instruments) to facilitate $in$-$situ$ pressure changes at low temperatures.
A designer diamond anvil (with a 180 $\mu$m culet size) equipped with tungsten leads was utilized~\cite{Weir_DesignerAnvil_2000}.
Solid steatite (soapstone) served as the insulating and pressure-transmitting medium on the stainless steel gasket.
Due to air sensitivity, the sample was loaded inside a nitrogen-filled glove box.
The cell screws were then used to secure the sample and apply pressure (\SI{1}{GPa}).
After attaching the membrane to the cell outside of the glove box, a small amount of helium gas pressure ($\sim$1 bar) was added to the membrane, and then the screws were removed.
Subsequently, the membrane was employed to further increase the pressure.
The sample was first pressurized at room temperature to \SI{2}{GPa} and any subsequent increases in pressure were performed below \SI{10}{K}.
Pressure was determined using the fluorescence of the $R_1$ peak of small ruby chips~\cite{chijioke_ruby_2005} or the Raman signal of the anvil~\cite{Akahama_RamanDiamondAnvil_2006}.

Resistance was measured in a four-probe arrangement using Keithly 6221 (DC current source) and Keithley 2182a (nanovoltmeter) configured for ``delta mode.''
The instruments were configured for resistance mode rather than voltage mode.
In resistance mode, the instrument reports absolute values, such that when the signal becomes small, the noise appears with a floor at zero (see the inset to Fig.~\ref{fig:fig1}).
The electrical resistivity was estimated using the van der Pauw method, assuming isotropy in the sample plane: ${\rho} = {\pi}tR/\ln{2}$, where $t$ represents the sample thickness ($\sim$10 $\mu$m).
The resistivity is accurate to roughly a factor of two or three, considering uncertainties in the initial thickness of the sample.
No adjustments were made for changes in sample thickness with pressure on the resistivity estimate.
Further details of the nonhydrostatic high-pressure resistivity technique, including a photograph using a designer diamond anvil, are provided in Ref.~\cite{Lim_Be22Re_2021}.

High-pressure and low-temperature X-ray diffraction (XRD) measurements were conducted using two symmetric diamond anvil cells (DACs) with a diamond culet size of 300 $\mu$m.
This setup was housed in a cryostat equipped with a double membrane-driven pressurizing system at beamline 16-BM-D, Advanced Photon Source, Argonne National Laboratory.
A 5N high-purity \ch{Sr} sample was loaded into the DACs with a 140 $\mu$m hole of pre-indented \ch{Re} gasket (from 250 $\mu$m down to 50 $\mu$m) inside an argon-filled glove box with \ch{O2} levels below \SI{0.5}{ppm} to prevent any oxidation, reaching initial pressures of 2-\SI{5}{GPa} without any pressure medium.
X-ray beams with wavelengths of $\SI{0.4133}{\angstrom}$ ($\SI{30}{\kilo\electronvolt}$) in Run 1 and $\SI{0.3100}{\angstrom}$ ($\SI{40}{\kilo\electronvolt}$) in Run 2 were focused to a spot $\sim$6$\times$$\SI{5}{\mu\m^2}$ (FWHM) in the sample with a tail of $\sim$30$\times$$\SI{40}{\mu\m^2}$.
The diffracted intensity was recorded using a Pilatus3 X CdTe 1M detector calibrated with a standard \ch{CeO2}, with exposure times typically ranging from 60 to 120 seconds per image.
Pressure was determined using the equation of state of \ch{Au} grains loaded into the sample chamber, reflecting the temperature dependence within ranges of 15-\SI{296}{K} using Eq. (13) and Table 4 in Ref.~\cite{Holzapfel_AuEOSatLowT_2001}.
DIOPTAS~\cite{Prescher_Dioptas_2015} software converted 2D diffraction images to 1D diffraction patterns, subsequently analyzed by the Rietveld~\cite{Rietveld_Rietveld_1969} or Le Bail~\cite{LeBail_LeBaiL_1988} methods using the GSAS-II software~\cite{Toby_GSASII_2013}.
Materials Project~\cite{Jain_MaterialsProject_2013} was used to obtain the Crystallographic Information File (CIF) data and modify it for the different phases of the \ch{Sr} sample, the \ch{Re} gasket and the pressure marker \ch{Au}.

\section{Results}
\begin{figure}[b]
    \centering
    \includegraphics[width=\columnwidth]{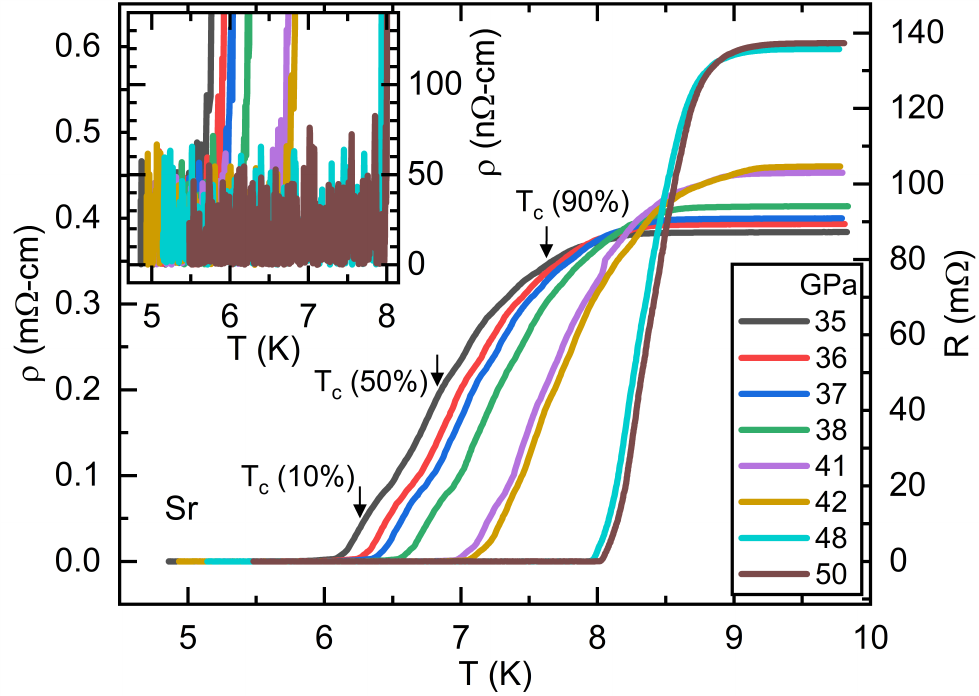}
    \caption{Temperature-dependent electrical resistivity of \ch{Sr} under pressure up to \SI{50}{GPa}, focusing on the superconducting transition.
    Pressure was applied below $\sim \SI{10}{K}$.
    Note that the transition becomes sharper above \SI{42}{GPa}, potentially indicating a phase transition. Three downward arrows represent the criteria for the superconducting critical temperature ($\Tc$) at \SI{35}{GPa} with $\Tc$(90\%), $\Tc$(50\%), and $\Tc$(10\%), respectively (see text). The inset shows a magnified view of the zero-resistivity with a mean of $\sim$20 n$\Omega$-cm. This figure is adapted from Ref.~\cite{Jackson_Thesis_2018}.}
    \label{fig:fig1}
\end{figure}

\begin{figure*}[t]
    \centering
    \includegraphics[width=\textwidth]{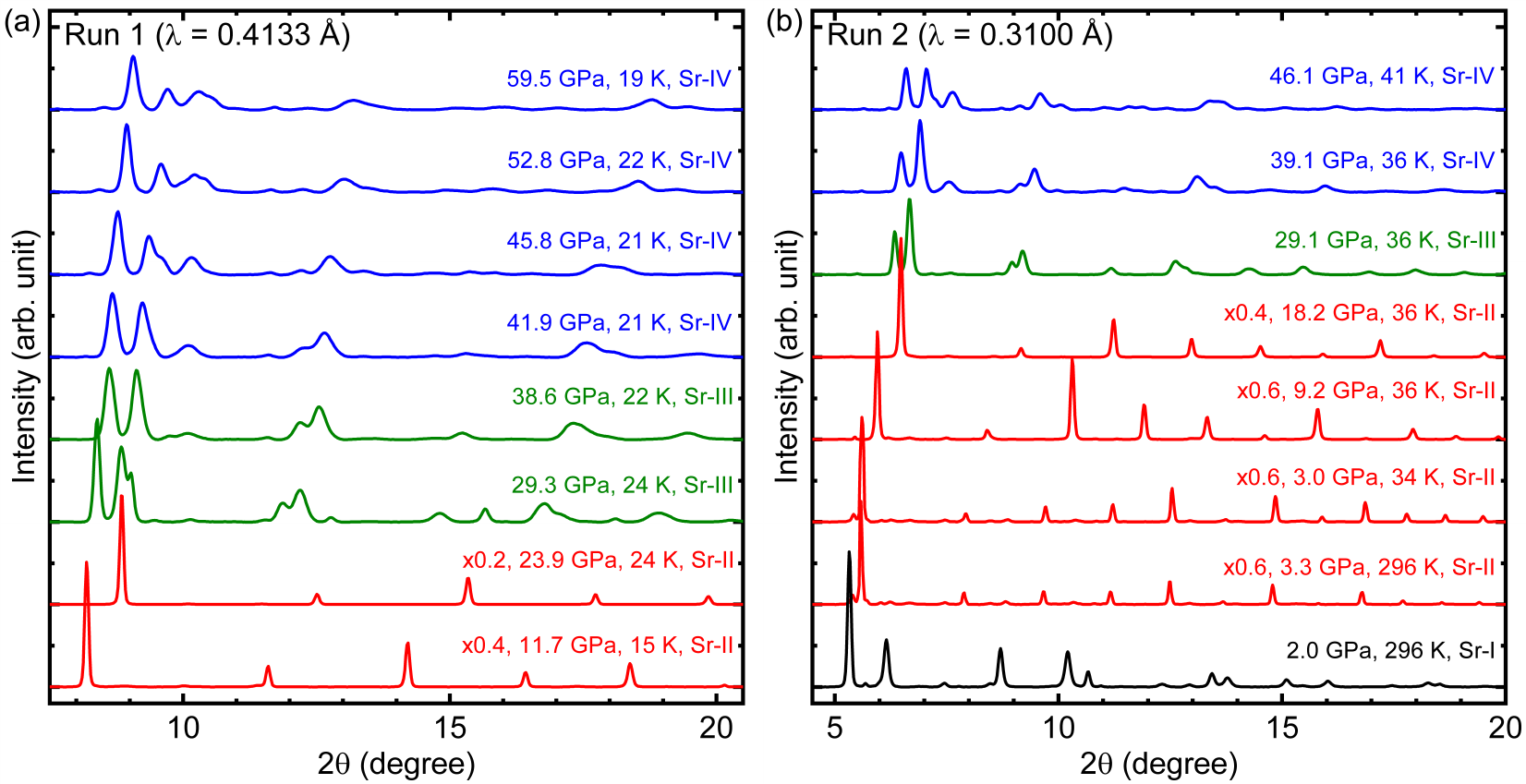}
    \caption{Representative high-pressure XRD patterns of \ch{Sr} under pressures up to \SI{60}{GPa} at low temperatures between 15-\SI{41}{K} (except for \SI{2.0}{GPa} and \SI{3.3}{GPa} at room temperature), showing the structural transition (a) from Sr-II (bcc) through Sr-III ($\beta$-tin) to Sr-IV (monoclinic) phases in Run 1 and (b) Sr-I (fcc) through Sr-II (bcc), Sr-III ($\beta$-tin) to Sr-IV (monoclinic) in Run 2. The diffraction profiles from Sr-III and Sr-IV are similar as they only differ in slight distortion~\cite{Katzke_BaSrHostGuest_2007}. The small peak around 5.7 degrees at \SI{2.0}{GPa} in Run 2 is unknown. A negligible second phase from \ch{Re} (hcp) gasket or potentially oxidized \ch{SrO} (fcc, $Fm\Bar{3}m$) or possibly the $S$ phase (previously reported as unknown) is present~\cite{Anzellini_ReEOS_2014, Liu_SrOfcc_1972, Allan_SrIII_1998, Bovornratanarak_SrIV_2006}.}
    \label{fig:fig2}
\end{figure*}

Figure~\ref{fig:fig1} displays the temperature-dependent electrical resistivity data of \ch{Sr} taken while loading pressure at low temperatures below \SI{10}{K}.
The superconducting critical temperature ($\Tc$) is defined as $\Tc$(90\%), $\Tc$(50\%), and $\Tc$(10\%) at which the resistivity has dropped to 90\%, 50\%, and 10\% of the normal-state resistivity just above the transition.
Superconductivity first appears at \SI{35}{GPa} with a broad transition, with $\Tc$(50\%) near \SI{7}{K} as shown with a downward arrow.
The superconducting transition width ($\Delta\Tc$) is defined as the difference between $\Tc$(90\%) and $\Tc$(10\%), which is depicted as the upper and lower vertical bars in Fig.~\ref{fig:fig4}(b)).
With increasing pressure, the transition temperature rises slowly accompanied by gradual increases in the normal-state resistivity, shown at \SI{9.7}{K} at each pressure.
When the pressure was increased to \SI{42}{GPa}, it suddenly jumped to \SI{48}{GPa}, although the resistivity in the normal state follows the same slope, increasing linearly with pressure.
Interestingly, the superconducting transition suddenly becomes significantly sharper above \SI{42}{GPa} to \SI{50}{GPa}, the maximum pressure reached, with $\Tc$(90\%) reaching \SI{8.7}{K} (see the upper and lower vertical bars, $\Delta\Tc$, of this work in Fig.~\ref{fig:fig4}(b)).
This abrupt change in $\Delta\Tc$ is reminiscent of the phase II to VI transition of \ch{Ba}, an isoelectronic element of Sr, at low temperatures below \SI{150}{K}~\cite{Desgreniers_BaStructure_2015}, as there is a sharpening of the superconducting transition above \SI{20}{GPa} across the structural transition~\cite{Jackson_BaSC_2017}.
It is not clear from the electrical resistivity data what may be happening in the pressure range across \SI{42}{GPa}.

To investigate the abrupt change in the $\Delta\Tc$ above \SI{42}{GPa}, we performed high-pressure and low-temperature XRD measurements on \ch{Sr}, as illustrated in Fig.~\ref{fig:fig2}.
The 1D XRD patterns shown are from the measurements performed at the center of the sample.
It is worth noting the presence of peak broadening with increasing pressure, indicating strain caused by non-hydrostatic pressure conditions.
At room temperature, a structural transition from Sr-I (fcc, $Fm\bar{3}m$) to Sr-II (bcc, $Im\bar{3}m$) occurs \SI{3.3}{GPa} in Run 2, depicted in Fig.~\ref{fig:fig2}(b), consistent with previous studies~\cite{McWhan_bccSr_1963}.
After cooling to temperatures between 15 and \SI{41}{K}, pressure was applied.
Above \SI{28}{GPa} at the low temperatures, Sr-II undergoes a transition to Sr-III ($\beta$-tin, $I4_1/amd$), slightly higher than the \SI{26}{GPa} reported at room temperature~\cite{Olijnyk_SrIII_1984, Allan_SrIII_1998}.
Sr-III persists up to \SI{39}{GPa}, where it transforms to Sr-IV (monoclinic, $Ia$).
The transition pressure at room temperature was reported to be \SI{35}{GPa}~\cite{Bovornratanarak_SrIV_2006}.
At low temperatures, Sr-IV remains stable up to \SI{59.5}{GPa}, the highest pressure measured, in contrast to the phase transition from Sr-IV to Sr-V (incommensurate, host-$I4/mcm$, guest-fct) at \SI{46}{GPa} at room temperature~\cite{McMahon_SrV_2000}.
This highlights the strong stability of Sr-IV at low temperatures compared to room temperature, ruling out the potential presence of a metastable structure in \ch{Sr}, unlike the metastable Ba-VI discovered only under low-temperature compression below the incommensurate structure of Ba-IV~\cite{Desgreniers_BaStructure_2015}.
The diffraction profiles of Sr-III and Sr-IV are similar, differing only in slight distortion~\cite{Katzke_BaSrHostGuest_2007}.

\begin{figure}[t]
    \centering
    \includegraphics[width=\columnwidth]{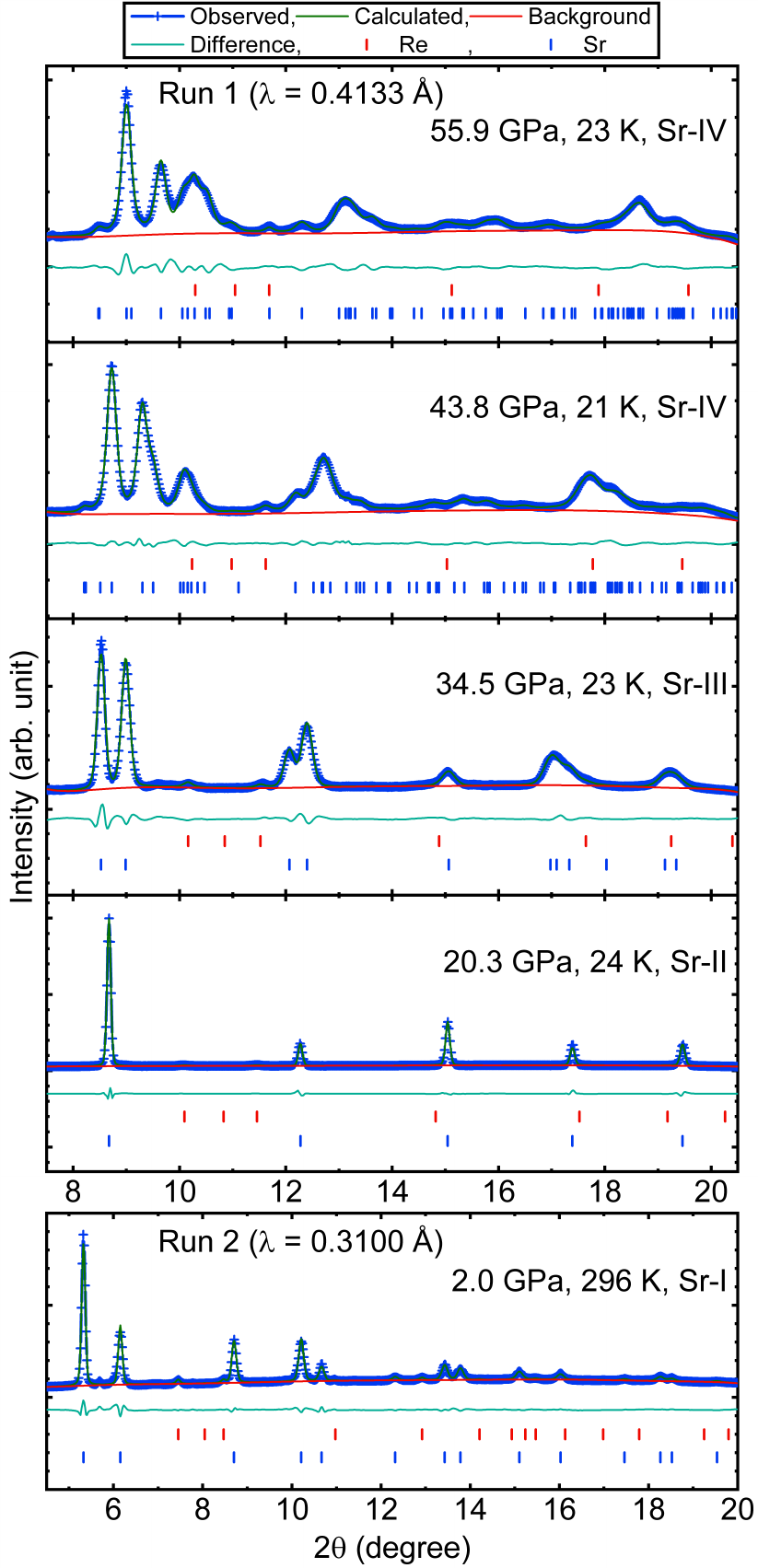}
    \caption{Representative Rietveld refinements of XRD patterns of \ch{Sr} at five different pressures from Runs 1 and 2, depicting Sr-I (fcc, $Fm\bar{3}m$), Sr-II (bcc, $Im\bar{3}m$), Sr-III ($\beta$-tin, $I4_1/amd$), and Sr-IV (monoclinic, $Ia$) phases. Note that the room-temperature phase Sr-V (incommensurate) is absent.}
    \label{fig:fig3}
\end{figure}

Figure~\ref{fig:fig3} shows representative Rietveld refinements of XRD patterns of \ch{Sr} at five different pressures, displaying various phases from Runs 1 and 2, including Sr-I phase at room temperature and Sr-II, Sr-III, Sr-IV phases at low temperatures.
The red and blue tick marks denote the expected locations of the Bragg peaks for the \ch{Re} gasket~\cite{Anzellini_ReEOS_2014} and \ch{Sr} samples, respectively.
A negligible second phase from the \ch{Re} gasket (hcp, $P6_{3}/mmc$)~\cite{Anzellini_ReEOS_2014} or \ch{SrO} (fcc, $Fm\Bar{3}m$)~\cite{Liu_SrOfcc_1972} is present.
It is noteworthy that the room-temperature phase Sr-V (incommensurate, host-$I4/mcm$, guest-fct)~\cite{McMahon_SrV_2000} is absent, which significantly differs in diffraction profiles from Sr-IV (monoclinic, $Ia$)~\cite{Bovornratanarak_SrIV_2006}.
All the refinements were carried out assuming preferred orientation of the crystalline grains using spherical harmonics since the raw detector images (see data repository~\cite{Zenodo_SrData_2024}) show slightly irregular intensity over the Debye-Scherrer rings (slightly textured).
All weighted profile $R$ factor ($\Rwp$), which provides the goodness-of-fit (GOF) estimates, are below 5\%. 
The XRD pattern at \SI{2.0}{GPa} at room temperature is well-matched with the fcc Sr-I phase with the space group $Fm\bar{3}m$ ($\Rwp$ = 4.084\%), in agreement with a previous study~\cite{McWhan_bccSr_1963}.
There is an unknown peak at around 5.7 degrees in 2$\theta$, which is different from both \ch{SrO} peaks and the 110 reflection (the first peak) from Sr-II arising in the phase boundary.
The XRD pattern at \SI{20.3}{GPa} and \SI{24}{K} is well-refined with the bcc Sr-II phase with the space group $Im\bar{3}m$ ($\Rwp$ = 4.075\%)~\cite{McWhan_bccSr_1963}.
The Sr-III phase is confirmed in the XRD pattern at \SI{34.5}{GPa} and \SI{23}{K}, which is the $\beta$-tin (bct) structure with the space group $I4_1/amd$ ($\Rwp$ = 4.010\%)~\cite{Olijnyk_SrIII_1984, Allan_SrIII_1998}.
At \SI{43.8}{GPa} and \SI{21}{K}, the diffraction pattern is refined with the monoclinic Sr-IV with the space group $Ia$ ($\Rwp$ = 2.017\%)~\cite{Bovornratanarak_SrIV_2006}.
The diffraction profiles of tetragonal $\beta$-tin Sr-III and monoclinic Sr-IV are similar because they differ only in slight distortion resulting in a helical chain in the Sr-IV phase along the previous Sr-III $c$-axis~\cite{Katzke_BaSrHostGuest_2007}.
This phase persists to higher pressure at \SI{55.9}{GPa} and \SI{23}{K} ($\Rwp$ = 3.5\%) without undergoing the transition to the incommensurate Sr-V~\cite{McMahon_SrV_2000}, which appears at room temperature.

\section{Discussion}
\begin{figure}[t]
    \centering
    \includegraphics[width=\columnwidth]{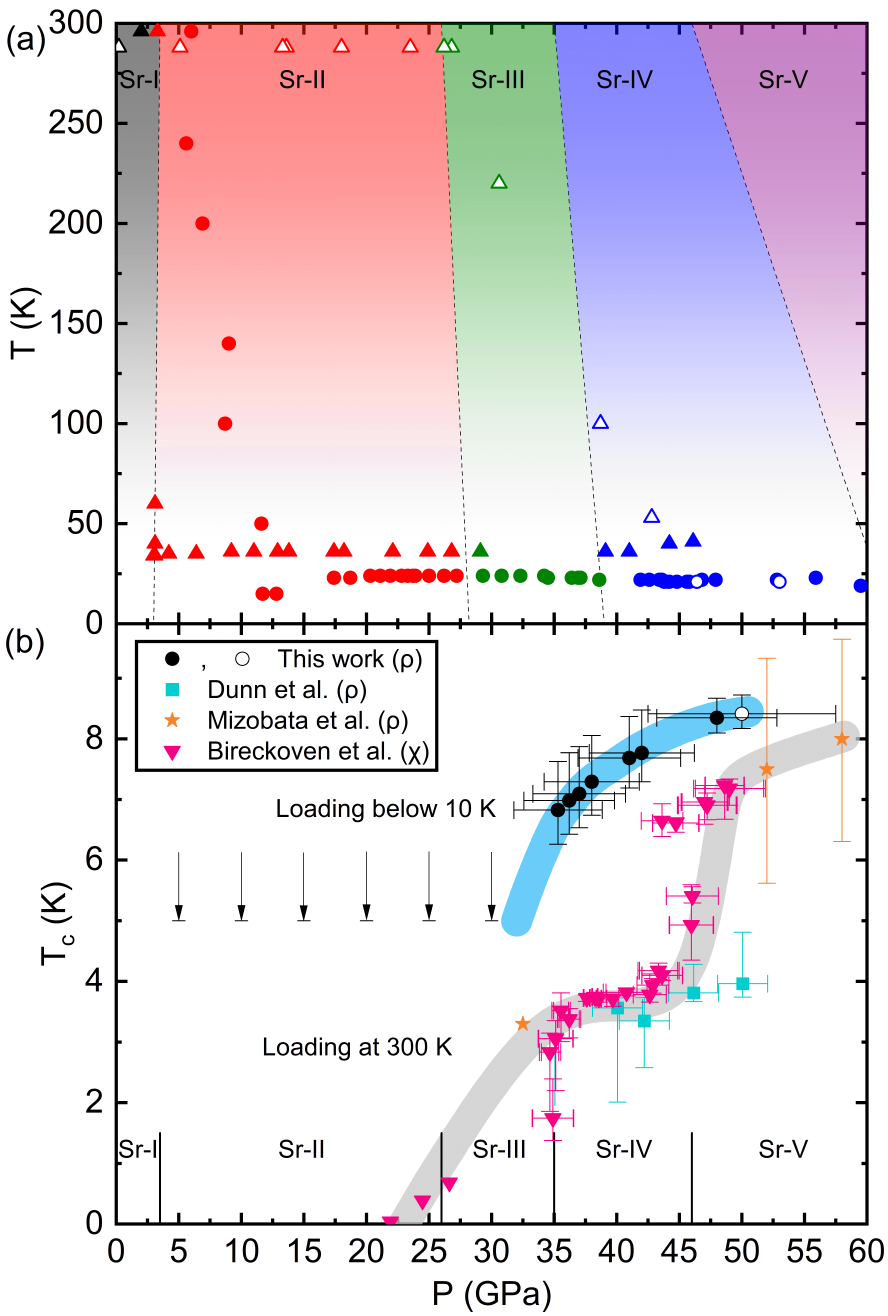}
    \caption{Phase diagram of \ch{Sr} focusing on (a) structural transitions and (b) corresponding superconductivity with the criterion $\Tc$(50\%) (see text). Sr-IV at low temperatures (blue area) exhibits strong phase stability covering a wide pressure range. Closed symbols (or open symbols) in (a) refer to data with loading pressure (or unloading pressure). A white circle in (b) indicates that the pressure was extrapolated from the relationship between membrane pressure (bar) and sample pressure (GPa). Black downward arrows with bars refer to the lowest temperatures (\SI{5}{K}) found with no superconductivity at each pressure point. $\rho$ and $\chi$ indicate electrical resistivity and magnetic susceptibility methods, respectively. The indications of the \ch{Sr} structure at room temperature are shown at the top and bottom of (a) and (b), respectively, taken from Ref.~\cite{McWhan_bccSr_1963, Olijnyk_SrIII_1984, Allan_SrIII_1998, Bovornratanarak_SrIV_2006, McMahon_SrV_2000}. The wide blue line in (b) corresponds to data taken with pressure applied at \SI{10}{K}, while the wide grey line corresponds to data where pressure was applied at room temperature. This figure is modified from Ref.~\cite{Jackson_Thesis_2018}.}
    \label{fig:fig4}
\end{figure}

Figure~\ref{fig:fig4} shows the phase diagrams of \ch{Sr} illustrating (a) structural transitions and (b) corresponding superconductivity, which are modified from Ref.~\cite{Jackson_Thesis_2018}.
The structural phase diagram covering 0-\SI{300}{K} and 0-\SI{60}{GPa} in Fig.~\ref{fig:fig4}(a) is constructed from the XRD data, as shown in Figs.~\ref{fig:fig2} and~\ref{fig:fig3}.
Run 1 (circle symbols) comprises 22 data points measured in the center of the sample and 24 data points measured in the position between the sample and the pressure marker \ch{Au} near the edge of \ch{Re} gasket.
Run 2 (triangle symbols) has 19 data points measured in the center and 14 data points measured in the mixed position of the sample and \ch{Au} located near the \ch{Re} gasket.
Black circles in Fig.~\ref{fig:fig4}(b) represent the $\Tc$(50\%) with loading pressure below \SI{10}{K}.
A white circle indicates that the pressure was extrapolated from the relationship between membrane pressure (bar) and sample pressure (GPa).
The upper and lower vertical bars  denote $\Tc$(90\%) and $\Tc$(10\%), respectively, indicating the transition width, $\Delta\Tc$.
The horizontal bars indicate the uncertainty of the pressure determination.
The indications of the \ch{Sr} structure at room temperature are shown at the top and bottom of (a) and (b), respectively, taken from Ref.~\cite{McWhan_bccSr_1963, Olijnyk_SrIII_1984, Allan_SrIII_1998, Bovornratanarak_SrIV_2006, McMahon_SrV_2000}.
The thick light blue and gray curves guide the eye for the data taken while compressing below \SI{10}{K} and previous measurements where pressure was varied at room temperature~\cite{Dunn_SrBaSC_1982, Mizobata_SrSC_2007, Bireckoven_SrBaSC_1989}, respectively.

The phase boundaries depicted in Fig.~\ref{fig:fig4}(a) shift towards higher pressures at low temperatures compared to room temperatures.
The bcc Sr-II phase occupies a broad region of the structural phase diagram up to \SI{28}{GPa} at low temperatures, where it transitions into the tetragonal $\beta$-tin Sr-III phase.
Notably, earlier theoretical studies~\cite{Phusittrakool_SrAbInitio_2008, Srepusharawoot_Sphase_2012, Kim_SrTheory_2012} suggested that the Sr-III phase is energetically unfavorable relative to others.
However, recent molecular dynamics (MD) and density functional theory (DFT) calculations by Tsuppayakorn-aek $et$ $al.$~\cite{Tsuppayakorn-aek_SrIIIMD_2015} have accurately reproduced the experimental structure sequence using the screened exchange local density approximation (sX-LDA) functional, which treats $d$ electrons differently compared to other functionals.
The upper boundary of the monoclinic Sr-IV phase in Fig.~\ref{fig:fig4}(a) extends significantly to much higher pressures, at least up to \SI{60}{GPa}, indicating its strong stability at low temperatures without transitioning to the incommensurate Sr-V structure present above \SI{46}{GPa} at room temperatures.
$Ab$ $initio$ random structure searching methods (AIRSS), coupled with density functional theory calculations by Kim $et$ $al.$~\cite{Kim_SrTheory_2012}, reveal that the Sr-IV phase has a formation enthalpy comparable to the Sr-V phase above \SI{35}{GPa}, with only a $\sim$\SI{16}{meV/atom} difference at \SI{50}{GPa} and zero temperature.
The absence of the incommensurate Sr-V phase at low temperatures suggests that thermal activation energy might be crucial in overcoming the kinetic barrier between the Sr-IV and Sr-V phases at room temperature~\cite{Lim_A15HP_2021}.
It is conceivable that low-temperature compression could mitigate the influence of unusually large anharmonic effects from the $d$ states near the Fermi level in \ch{Sr}~\cite{Vaks_SrAnharmonic_1988, Jackson_Thesis_2018}, thereby reinforcing the strong phase stability of the monoclinic Sr-IV phase.
The XRD measurements during unloading (depicted as open circle or triangle symbols) in Fig.~\ref{fig:fig4}(a) demonstrate the reversibility of the structural phase boundaries.

Figure~\ref{fig:fig4}(b) illustrates the superconducting phase diagram of \ch{Sr}, including samples compressed at low temperatures from the current study (Fig.~\ref{fig:fig1} and those pressurized at room temperatures from previous studies~\cite{Dunn_SrBaSC_1982, Mizobata_SrSC_2007, Bireckoven_SrBaSC_1989}).
A comparison to previously studied samples by Dunn $et$ $al.$~\cite{Dunn_SrBaSC_1982} and Bireckoven $et$ $al.$~\cite{Bireckoven_SrBaSC_1989} within the Sr-IV phase pressure range at room temperature reveals that \ch{Sr} compressed at low temperatures exhibits a $\Tc$ approximately twice as high.
This observation could imply the presence of a structure different from Sr-IV when compressed at low temperatures, possibly indicating a metastable structure.
However, the present XRD measurements find no additional structure at low-temperature compression to \SI{60}{GPa}.
Rather, they indicate that applying pressure at low temperatures stabilizes the Sr-III phase to a higher pressure, reaching \SI{39}{GPa}, overlapping with the Sr-IV phase at room temperature, as depicted in Fig.~\ref{fig:fig4}(a).
This suggests that the Sr-III structure is being quenched to higher pressure at low temperatures, similar to the other phases.
Notably, the Sr (III-IV) transition above \SI{39}{GPa} at low temperatures aligns closely with the pressure where the change in the width $\Delta\Tc$ occurs above \SI{42}{GPa} (within the small uncertainty of pressure determination).
Additionally, a positive slope of $d\Tc/dP$ in the Sr-III phase compressed at low temperature corresponds to that in the Sr-III phase compressed at room temperature.

The most recent electrical resistivity measurements by Mizobata $et$ $al.$~\cite{Mizobata_SrSC_2007} closely align with the $\Tc(P)$ dependencies observed in the present study, as depicted in Fig.~\ref{fig:fig4}(b), due to similar experimental conditions.
Both studies involved loading high-purity samples in a glove box and utilizing non-hydrostatic pressure environments in DACs.
Minor variations in the $\Tc(P)$ data likely arise from pressure gradients across the samples and shear stress induced by the non-hydrostatic pressure medium.
It is widely recognized that such shear stresses can significantly influence $\Tc(P)$ dependencies~\cite{Schilling_AlkaliHP_2006}.
Note that some of the $\Tc(P)$ data above \SI{42}{GPa}, measured by Bireckoven $et$ $al.$~\cite{Bireckoven_SrBaSC_1989}, agree with the present study, where it turns to the monoclinic Sr-IV at low temperatures.
Kim \textit{et al.}~\cite{Kim_SrTheory_2012}, employing AIRSS, DFT, electron-phonon coupling calculations, investigated the phase stability and superconductivity of \ch{Sr} up to \SI{50}{GPa} at \SI{0}{K}.
They identified an orthorhombic structure ($Cmcm$) as stable above \SI{25}{GPa}, exhibiting a $\Tc$ of around \SI{4}{K} within the pressure range of 30-\SI{55}{GPa}.
This finding corresponds well the experimental $\Tc$ measured by Dunn $et$ $al.$~\cite{Dunn_SrBaSC_1982} and Bireckoven $et$ $al.$~\cite{Bireckoven_SrBaSC_1989}, as shown in Fig.~\ref{fig:fig4}(b).
However, the discrepancy in $\Tc(P)$ data between low-temperature and room-temperature compression, particularly in the pressure range of 35-\SI{50}{GPa}, remains unclear and does not seem to stem from any potential metastable structure according to the current low-temperature XRD measurements.
Dunn $et$ $al.$~\cite{Dunn_SrBaSC_1982} reported that the surface of the measured sample (99\% purity) was oxidized, leading to a significant initial contact resistance.
The presence of contamination could lead to sample homogeneity and suppress $\Tc$~\cite{Lim_NbMoB2_2023}.

\section{Conclusions}
In summary, we conducted measurements on the superconducting and structural properties of elemental \ch{Sr} metal up to \SI{60}{GPa} under low-temperature compression.
We observed a twofold increase in $\Tc$ between 35-\SI{42}{GPa} compared to cryogenic cooling after room-temperature compression.
Subsequently, the transition width became significantly sharper above \SI{42}{GPa}.
The results of low-temperature X-ray diffraction measurements under pressure indicate that this change corresponds to the Sr-III to Sr-IV) transition, without the presence of any potential metastable structure similar to the high-pressure phase that was observed in cryogenic compression experiments on elemental Ba~\cite{Loa_BIVc_2012,Jackson_BaSC_2017}.
Additionally, the monoclinic Sr-IV structure exhibited stability at significantly elevated pressures, reaching up to \SI{60}{GPa}, without transitioning into the incommensurate Sr-V phase observed at room temperature.
This suggests that thermal activation energy plays a crucial role in overcoming the kinetic barrier to the Sr-V phase at ambient temperature.

\vspace{1em}
\section{Data availability}
All primary and analyzed data associated with the present study are publicly available via zenodo~\cite{Zenodo_SrData_2024}.

\section*{Acknowledgments}
The development of high-pressure equipment and collection of electrical resistance data at the University of Florida was supported by the National Science Foundation CAREER award (DMR-1453752).
Work at the University of Florida was performed under the auspices of the U.S. Department of Energy Basic Energy Sciences under Contract No.\ DE-SC-0020385 (xray data collection) and the U.S. National Science Foundation, Division of Materials Research under Contract No.\ NSF-DMR-2118718.
We thank G.\ Fabbris (X-ray Science Division, Argonne National Laboratory) for his help in determining the pressure of the temperature-dependent equation of state of \ch{Au}, and C. Kenney-Benson and R.\ Ferry for technical assistance (HPCAT).
R.S.K.\ and R.J.H.\ acknowledge support from the U.S.\ National Science Foundation (DMR-2119308 and DMR-2104881).
XRD measurements were performed at HPCAT (Sector 16), Advanced Photon Source (APS), Argonne National Laboratory.
HPCAT operations are supported by the DOE-National Nuclear Security Administration (NNSA) Office of Experimental Sciences.
The beamtime was made possible by the Chicago/DOE Alliance Center (CDAC), which is supported by DOE-NNSA (DE-NA0003975).
The Advanced Photon Source is a DOE Office of Science User Facility operated for the DOE Office of Science by Argonne National Laboratory under Contract No. DE-AC02-06CH11357.


\begin{thebibliography}{41}%
\makeatletter
\providecommand \@ifxundefined [1]{%
 \@ifx{#1\undefined}
}%
\providecommand \@ifnum [1]{%
 \ifnum #1\expandafter \@firstoftwo
 \else \expandafter \@secondoftwo
 \fi
}%
\providecommand \@ifx [1]{%
 \ifx #1\expandafter \@firstoftwo
 \else \expandafter \@secondoftwo
 \fi
}%
\providecommand \natexlab [1]{#1}%
\providecommand \enquote  [1]{``#1''}%
\providecommand \bibnamefont  [1]{#1}%
\providecommand \bibfnamefont [1]{#1}%
\providecommand \citenamefont [1]{#1}%
\providecommand \href@noop [0]{\@secondoftwo}%
\providecommand \href [0]{\begingroup \@sanitize@url \@href}%
\providecommand \@href[1]{\@@startlink{#1}\@@href}%
\providecommand \@@href[1]{\endgroup#1\@@endlink}%
\providecommand \@sanitize@url [0]{\catcode `\\12\catcode `\$12\catcode
  `\&12\catcode `\#12\catcode `\^12\catcode `\_12\catcode `\%12\relax}%
\providecommand \@@startlink[1]{}%
\providecommand \@@endlink[0]{}%
\providecommand \url  [0]{\begingroup\@sanitize@url \@url }%
\providecommand \@url [1]{\endgroup\@href {#1}{\urlprefix }}%
\providecommand \urlprefix  [0]{URL }%
\providecommand \Eprint [0]{\href }%
\providecommand \doibase [0]{http://dx.doi.org/}%
\providecommand \selectlanguage [0]{\@gobble}%
\providecommand \bibinfo  [0]{\@secondoftwo}%
\providecommand \bibfield  [0]{\@secondoftwo}%
\providecommand \translation [1]{[#1]}%
\providecommand \BibitemOpen [0]{}%
\providecommand \bibitemStop [0]{}%
\providecommand \bibitemNoStop [0]{.\EOS\space}%
\providecommand \EOS [0]{\spacefactor3000\relax}%
\providecommand \BibitemShut  [1]{\csname bibitem#1\endcsname}%
\let\auto@bib@innerbib\@empty
\bibitem [{\citenamefont {Tse}(2019)}]{Tse_HPCristalReview_2020}%
  \BibitemOpen
  \bibfield  {author} {\bibinfo {author} {\bibfnamefont {J.~S.}\ \bibnamefont
  {Tse}},\ }\href {\doibase 10.1093/nsr/nwz144} {\bibfield  {journal} {\bibinfo
   {journal} {National Science Review}\ }\textbf {\bibinfo {volume} {7}},\
  \bibinfo {pages} {149} (\bibinfo {year} {2019})}\BibitemShut {NoStop}%
\bibitem [{\citenamefont {Cannon}(1974)}]{Cannon_ElementReview_1974}%
  \BibitemOpen
  \bibfield  {author} {\bibinfo {author} {\bibfnamefont {J.~F.}\ \bibnamefont
  {Cannon}},\ }\href {\doibase 10.1063/1.3253148} {\bibfield  {journal}
  {\bibinfo  {journal} {Journal of Physical and Chemical Reference Data}\
  }\textbf {\bibinfo {volume} {3}},\ \bibinfo {pages} {781} (\bibinfo {year}
  {1974})}\BibitemShut {NoStop}%
\bibitem [{\citenamefont {Ackland}\ \emph {et~al.}(2017)\citenamefont
  {Ackland}, \citenamefont {Dunuwille}, \citenamefont {Martinez-Canales},
  \citenamefont {Loa}, \citenamefont {Zhang}, \citenamefont {Sinogeikin},
  \citenamefont {Cai},\ and\ \citenamefont {Deemyad}}]{Ackland_Lifcc_2017}%
  \BibitemOpen
  \bibfield  {author} {\bibinfo {author} {\bibfnamefont {G.~J.}\ \bibnamefont
  {Ackland}}, \bibinfo {author} {\bibfnamefont {M.}~\bibnamefont {Dunuwille}},
  \bibinfo {author} {\bibfnamefont {M.}~\bibnamefont {Martinez-Canales}},
  \bibinfo {author} {\bibfnamefont {I.}~\bibnamefont {Loa}}, \bibinfo {author}
  {\bibfnamefont {R.}~\bibnamefont {Zhang}}, \bibinfo {author} {\bibfnamefont
  {S.}~\bibnamefont {Sinogeikin}}, \bibinfo {author} {\bibfnamefont
  {W.}~\bibnamefont {Cai}}, \ and\ \bibinfo {author} {\bibfnamefont
  {S.}~\bibnamefont {Deemyad}},\ }\href {\doibase 10.1126/science.aal4886}
  {\bibfield  {journal} {\bibinfo  {journal} {Science}\ }\textbf {\bibinfo
  {volume} {356}},\ \bibinfo {pages} {1254} (\bibinfo {year}
  {2017})}\BibitemShut {NoStop}%
\bibitem [{\citenamefont {Desgreniers}\ \emph {et~al.}(2015)\citenamefont
  {Desgreniers}, \citenamefont {Tse}, \citenamefont {Matsuoka}, \citenamefont
  {Ohishi}, \citenamefont {Li},\ and\ \citenamefont
  {Ma}}]{Desgreniers_BaStructure_2015}%
  \BibitemOpen
  \bibfield  {author} {\bibinfo {author} {\bibfnamefont {S.}~\bibnamefont
  {Desgreniers}}, \bibinfo {author} {\bibfnamefont {J.~S.}\ \bibnamefont
  {Tse}}, \bibinfo {author} {\bibfnamefont {T.}~\bibnamefont {Matsuoka}},
  \bibinfo {author} {\bibfnamefont {Y.}~\bibnamefont {Ohishi}}, \bibinfo
  {author} {\bibfnamefont {Q.}~\bibnamefont {Li}}, \ and\ \bibinfo {author}
  {\bibfnamefont {Y.}~\bibnamefont {Ma}},\ }\href
  {https://doi.org/10.1063/1.4936849} {\bibfield  {journal} {\bibinfo
  {journal} {Applied Physics Letters}\ }\textbf {\bibinfo {volume} {107}},\
  \bibinfo {pages} {221908} (\bibinfo {year} {2015})}\BibitemShut {NoStop}%
\bibitem [{\citenamefont {Jackson}\ \emph {et~al.}(2017)\citenamefont
  {Jackson}, \citenamefont {VanGennep}, \citenamefont {Vohra}, \citenamefont
  {Weir},\ and\ \citenamefont {Hamlin}}]{Jackson_BaSC_2017}%
  \BibitemOpen
  \bibfield  {author} {\bibinfo {author} {\bibfnamefont {D.~E.}\ \bibnamefont
  {Jackson}}, \bibinfo {author} {\bibfnamefont {D.}~\bibnamefont {VanGennep}},
  \bibinfo {author} {\bibfnamefont {Y.~K.}\ \bibnamefont {Vohra}}, \bibinfo
  {author} {\bibfnamefont {S.~T.}\ \bibnamefont {Weir}}, \ and\ \bibinfo
  {author} {\bibfnamefont {J.~J.}\ \bibnamefont {Hamlin}},\ }\href {\doibase
  10.1103/PhysRevB.96.184514} {\bibfield  {journal} {\bibinfo  {journal} {Phys.
  Rev. B}\ }\textbf {\bibinfo {volume} {96}},\ \bibinfo {pages} {184514}
  (\bibinfo {year} {2017})}\BibitemShut {NoStop}%
\bibitem [{\citenamefont {McMahon}\ \emph {et~al.}(2000)\citenamefont
  {McMahon}, \citenamefont {Bovornratanaraks}, \citenamefont {Allan},
  \citenamefont {Belmonte},\ and\ \citenamefont {Nelmes}}]{McMahon_SrV_2000}%
  \BibitemOpen
  \bibfield  {author} {\bibinfo {author} {\bibfnamefont {M.~I.}\ \bibnamefont
  {McMahon}}, \bibinfo {author} {\bibfnamefont {T.}~\bibnamefont
  {Bovornratanaraks}}, \bibinfo {author} {\bibfnamefont {D.~R.}\ \bibnamefont
  {Allan}}, \bibinfo {author} {\bibfnamefont {S.~A.}\ \bibnamefont {Belmonte}},
  \ and\ \bibinfo {author} {\bibfnamefont {R.~J.}\ \bibnamefont {Nelmes}},\
  }\href {\doibase 10.1103/PhysRevB.61.3135} {\bibfield  {journal} {\bibinfo
  {journal} {Phys. Rev. B}\ }\textbf {\bibinfo {volume} {61}},\ \bibinfo
  {pages} {3135} (\bibinfo {year} {2000})}\BibitemShut {NoStop}%
\bibitem [{\citenamefont {Nelmes}\ \emph {et~al.}(1999)\citenamefont {Nelmes},
  \citenamefont {Allan}, \citenamefont {McMahon},\ and\ \citenamefont
  {Belmonte}}]{Nelmes_BaIV_1999}%
  \BibitemOpen
  \bibfield  {author} {\bibinfo {author} {\bibfnamefont {R.~J.}\ \bibnamefont
  {Nelmes}}, \bibinfo {author} {\bibfnamefont {D.~R.}\ \bibnamefont {Allan}},
  \bibinfo {author} {\bibfnamefont {M.~I.}\ \bibnamefont {McMahon}}, \ and\
  \bibinfo {author} {\bibfnamefont {S.~A.}\ \bibnamefont {Belmonte}},\ }\href
  {\doibase 10.1103/PhysRevLett.83.4081} {\bibfield  {journal} {\bibinfo
  {journal} {Phys. Rev. Lett.}\ }\textbf {\bibinfo {volume} {83}},\ \bibinfo
  {pages} {4081} (\bibinfo {year} {1999})}\BibitemShut {NoStop}%
\bibitem [{\citenamefont {Loa}\ \emph {et~al.}(2012)\citenamefont {Loa},
  \citenamefont {Nelmes}, \citenamefont {Lundegaard},\ and\ \citenamefont
  {McMahon}}]{Loa_BIVc_2012}%
  \BibitemOpen
  \bibfield  {author} {\bibinfo {author} {\bibfnamefont {I.}~\bibnamefont
  {Loa}}, \bibinfo {author} {\bibfnamefont {R.~J.}\ \bibnamefont {Nelmes}},
  \bibinfo {author} {\bibfnamefont {L.~F.}\ \bibnamefont {Lundegaard}}, \ and\
  \bibinfo {author} {\bibfnamefont {M.~I.}\ \bibnamefont {McMahon}},\ }\href
  {\doibase 10.1038/nmat3342} {\bibfield  {journal} {\bibinfo  {journal}
  {Nature Materials}\ }\textbf {\bibinfo {volume} {11}},\ \bibinfo {pages}
  {627} (\bibinfo {year} {2012})}\BibitemShut {NoStop}%
\bibitem [{\citenamefont {McWhan}\ and\ \citenamefont
  {Jayaraman}(1963)}]{McWhan_bccSr_1963}%
  \BibitemOpen
  \bibfield  {author} {\bibinfo {author} {\bibfnamefont {D.~B.}\ \bibnamefont
  {McWhan}}\ and\ \bibinfo {author} {\bibfnamefont {A.}~\bibnamefont
  {Jayaraman}},\ }\href {https://doi.org/10.1063/1.1753899} {\bibfield
  {journal} {\bibinfo  {journal} {Applied Physics Letters}\ }\textbf {\bibinfo
  {volume} {3}},\ \bibinfo {pages} {129} (\bibinfo {year} {1963})}\BibitemShut
  {NoStop}%
\bibitem [{\citenamefont {Olijnyk}\ and\ \citenamefont
  {Holzapfel}(1984)}]{Olijnyk_SrIII_1984}%
  \BibitemOpen
  \bibfield  {author} {\bibinfo {author} {\bibfnamefont {H.}~\bibnamefont
  {Olijnyk}}\ and\ \bibinfo {author} {\bibfnamefont {W.}~\bibnamefont
  {Holzapfel}},\ }\href {\doibase https://doi.org/10.1016/0375-9601(84)90757-6}
  {\bibfield  {journal} {\bibinfo  {journal} {Physics Letters A}\ }\textbf
  {\bibinfo {volume} {100}},\ \bibinfo {pages} {191} (\bibinfo {year}
  {1984})}\BibitemShut {NoStop}%
\bibitem [{\citenamefont {Allan}\ \emph {et~al.}(1998)\citenamefont {Allan},
  \citenamefont {Nelmes}, \citenamefont {McMahon}, \citenamefont {Belmonte},\
  and\ \citenamefont {Bovornratanaraks}}]{Allan_SrIII_1998}%
  \BibitemOpen
  \bibfield  {author} {\bibinfo {author} {\bibfnamefont {D.~R.}\ \bibnamefont
  {Allan}}, \bibinfo {author} {\bibfnamefont {R.~J.}\ \bibnamefont {Nelmes}},
  \bibinfo {author} {\bibfnamefont {M.~I.}\ \bibnamefont {McMahon}}, \bibinfo
  {author} {\bibfnamefont {S.~A.}\ \bibnamefont {Belmonte}}, \ and\ \bibinfo
  {author} {\bibfnamefont {T.}~\bibnamefont {Bovornratanaraks}},\ }\href
  {\doibase 10.4131/jshpreview.7.236} {\bibfield  {journal} {\bibinfo
  {journal} {Rev. High Pressure Sci. Technol.}\ }\textbf {\bibinfo {volume}
  {7}},\ \bibinfo {pages} {236} (\bibinfo {year} {1998})}\BibitemShut {NoStop}%
\bibitem [{\citenamefont {Bovornratanaraks}\ \emph {et~al.}(2006)\citenamefont
  {Bovornratanaraks}, \citenamefont {Allan}, \citenamefont {Belmonte},
  \citenamefont {McMahon},\ and\ \citenamefont
  {Nelmes}}]{Bovornratanarak_SrIV_2006}%
  \BibitemOpen
  \bibfield  {author} {\bibinfo {author} {\bibfnamefont {T.}~\bibnamefont
  {Bovornratanaraks}}, \bibinfo {author} {\bibfnamefont {D.~R.}\ \bibnamefont
  {Allan}}, \bibinfo {author} {\bibfnamefont {S.~A.}\ \bibnamefont {Belmonte}},
  \bibinfo {author} {\bibfnamefont {M.~I.}\ \bibnamefont {McMahon}}, \ and\
  \bibinfo {author} {\bibfnamefont {R.~J.}\ \bibnamefont {Nelmes}},\ }\href
  {\doibase 10.1103/PhysRevB.73.144112} {\bibfield  {journal} {\bibinfo
  {journal} {Phys. Rev. B}\ }\textbf {\bibinfo {volume} {73}},\ \bibinfo
  {pages} {144112} (\bibinfo {year} {2006})}\BibitemShut {NoStop}%
\bibitem [{\citenamefont {Katzke}\ and\ \citenamefont
  {Tol\'edano}(2007)}]{Katzke_BaSrHostGuest_2007}%
  \BibitemOpen
  \bibfield  {author} {\bibinfo {author} {\bibfnamefont {H.}~\bibnamefont
  {Katzke}}\ and\ \bibinfo {author} {\bibfnamefont {P.}~\bibnamefont
  {Tol\'edano}},\ }\href {\doibase 10.1103/PhysRevB.75.174103} {\bibfield
  {journal} {\bibinfo  {journal} {Phys. Rev. B}\ }\textbf {\bibinfo {volume}
  {75}},\ \bibinfo {pages} {174103} (\bibinfo {year} {2007})}\BibitemShut
  {NoStop}%
\bibitem [{\citenamefont {Jackson}(2018)}]{Jackson_Thesis_2018}%
  \BibitemOpen
  \bibfield  {author} {\bibinfo {author} {\bibfnamefont {D.~E.}\ \bibnamefont
  {Jackson}},\ }\href
  {https://ufl-flvc.primo.exlibrisgroup.com/permalink/01FALSC_UFL/175ga98/alma990366927560306597}
  {\bibfield  {journal} {\bibinfo  {journal} {Ph.D. thesis, University of
  Florida}\ } (\bibinfo {year} {2018})}\BibitemShut {NoStop}%
\bibitem [{\citenamefont {Dunn}\ and\ \citenamefont
  {Bundy}(1982)}]{Dunn_SrBaSC_1982}%
  \BibitemOpen
  \bibfield  {author} {\bibinfo {author} {\bibfnamefont {K.~J.}\ \bibnamefont
  {Dunn}}\ and\ \bibinfo {author} {\bibfnamefont {F.~P.}\ \bibnamefont
  {Bundy}},\ }\href {\doibase 10.1103/PhysRevB.25.194} {\bibfield  {journal}
  {\bibinfo  {journal} {Phys. Rev. B}\ }\textbf {\bibinfo {volume} {25}},\
  \bibinfo {pages} {194} (\bibinfo {year} {1982})}\BibitemShut {NoStop}%
\bibitem [{\citenamefont {Mizobata}\ \emph {et~al.}(2007)\citenamefont
  {Mizobata}, \citenamefont {Matsuoka},\ and\ \citenamefont
  {Shimizu}}]{Mizobata_SrSC_2007}%
  \BibitemOpen
  \bibfield  {author} {\bibinfo {author} {\bibfnamefont {S.}~\bibnamefont
  {Mizobata}}, \bibinfo {author} {\bibfnamefont {T.}~\bibnamefont {Matsuoka}},
  \ and\ \bibinfo {author} {\bibfnamefont {K.}~\bibnamefont {Shimizu}},\ }\href
  {\doibase 10.1143/JPSJS.76SA.23} {\bibfield  {journal} {\bibinfo  {journal}
  {Journal of the Physical Society of Japan}\ }\textbf {\bibinfo {volume}
  {76}},\ \bibinfo {pages} {23} (\bibinfo {year} {2007})}\BibitemShut {NoStop}%
\bibitem [{\citenamefont {Bireckoven}\ and\ \citenamefont
  {Wittig}()}]{Bireckoven_SrBaSC_1989}%
  \BibitemOpen
  \bibfield  {author} {\bibinfo {author} {\bibfnamefont {B.}~\bibnamefont
  {Bireckoven}}\ and\ \bibinfo {author} {\bibfnamefont {J.}~\bibnamefont
  {Wittig}},\ }\href {https://www.airapt.org/proceedings/} {\bibinfo  {journal}
  {in Proc. 11 Int. Conf. AIRAPT, High Pressure Science and Technology, Vol. 3
  (Naukova Dumka, 1989) pp. 14-24}\ }\BibitemShut {NoStop}%
\bibitem [{\citenamefont {Hamlin}(2015)}]{Hamlin_ElementsHP_2015}%
  \BibitemOpen
\bibfield  {journal} {  }\bibfield  {author} {\bibinfo {author} {\bibfnamefont
  {J.}~\bibnamefont {Hamlin}},\ }\href {\doibase
  https://doi.org/10.1016/j.physc.2015.02.032} {\bibfield  {journal} {\bibinfo
  {journal} {Physica C: Superconductivity and its Applications}\ }\textbf
  {\bibinfo {volume} {514}},\ \bibinfo {pages} {59} (\bibinfo {year}
  {2015})}\BibitemShut {NoStop}%
\bibitem [{\citenamefont {Skriver}(1982)}]{Skriver_Srndcal_1982}%
  \BibitemOpen
  \bibfield  {author} {\bibinfo {author} {\bibfnamefont {H.~L.}\ \bibnamefont
  {Skriver}},\ }\href {\doibase 10.1103/PhysRevLett.49.1768} {\bibfield
  {journal} {\bibinfo  {journal} {Phys. Rev. Lett.}\ }\textbf {\bibinfo
  {volume} {49}},\ \bibinfo {pages} {1768} (\bibinfo {year}
  {1982})}\BibitemShut {NoStop}%
\bibitem [{\citenamefont {Skriver}(1985)}]{Skriver_ndcalculations_1985}%
  \BibitemOpen
  \bibfield  {author} {\bibinfo {author} {\bibfnamefont {H.~L.}\ \bibnamefont
  {Skriver}},\ }\href {\doibase 10.1103/PhysRevB.31.1909} {\bibfield  {journal}
  {\bibinfo  {journal} {Phys. Rev. B}\ }\textbf {\bibinfo {volume} {31}},\
  \bibinfo {pages} {1909} (\bibinfo {year} {1985})}\BibitemShut {NoStop}%
\bibitem [{\citenamefont {Weir}\ \emph {et~al.}(2000)\citenamefont {Weir},
  \citenamefont {Akella}, \citenamefont {Aracne-Ruddle}, \citenamefont
  {Vohra},\ and\ \citenamefont {Catledge}}]{Weir_DesignerAnvil_2000}%
  \BibitemOpen
  \bibfield  {author} {\bibinfo {author} {\bibfnamefont {S.~T.}\ \bibnamefont
  {Weir}}, \bibinfo {author} {\bibfnamefont {J.}~\bibnamefont {Akella}},
  \bibinfo {author} {\bibfnamefont {C.}~\bibnamefont {Aracne-Ruddle}}, \bibinfo
  {author} {\bibfnamefont {Y.~K.}\ \bibnamefont {Vohra}}, \ and\ \bibinfo
  {author} {\bibfnamefont {S.~A.}\ \bibnamefont {Catledge}},\ }\href {\doibase
  10.1063/1.1326838} {\bibfield  {journal} {\bibinfo  {journal} {Applied
  Physics Letters}\ }\textbf {\bibinfo {volume} {77}},\ \bibinfo {pages} {3400}
  (\bibinfo {year} {2000})}\BibitemShut {NoStop}%
\bibitem [{\citenamefont {Chijioke}\ \emph {et~al.}(2005)\citenamefont
  {Chijioke}, \citenamefont {Nellis}, \citenamefont {Soldatov},\ and\
  \citenamefont {Silvera}}]{chijioke_ruby_2005}%
  \BibitemOpen
  \bibfield  {author} {\bibinfo {author} {\bibfnamefont {A.~D.}\ \bibnamefont
  {Chijioke}}, \bibinfo {author} {\bibfnamefont {W.~J.}\ \bibnamefont
  {Nellis}}, \bibinfo {author} {\bibfnamefont {A.}~\bibnamefont {Soldatov}}, \
  and\ \bibinfo {author} {\bibfnamefont {I.~F.}\ \bibnamefont {Silvera}},\
  }\href {\doibase 10.1063/1.2135877} {\bibfield  {journal} {\bibinfo
  {journal} {Journal of Applied Physics}\ }\textbf {\bibinfo {volume} {98}},\
  \bibinfo {pages} {114905} (\bibinfo {year} {2005})}\BibitemShut {NoStop}%
\bibitem [{\citenamefont {Akahama}\ and\ \citenamefont
  {Kawamura}(2006)}]{Akahama_RamanDiamondAnvil_2006}%
  \BibitemOpen
  \bibfield  {author} {\bibinfo {author} {\bibfnamefont {Y.}~\bibnamefont
  {Akahama}}\ and\ \bibinfo {author} {\bibfnamefont {H.}~\bibnamefont
  {Kawamura}},\ }\href {\doibase 10.1063/1.2335683} {\bibfield  {journal}
  {\bibinfo  {journal} {Journal of Applied Physics}\ }\textbf {\bibinfo
  {volume} {100}},\ \bibinfo {pages} {043516} (\bibinfo {year}
  {2006})}\BibitemShut {NoStop}%
\bibitem [{\citenamefont {Lim}\ \emph {et~al.}(2021{\natexlab{a}})\citenamefont
  {Lim}, \citenamefont {Hire}, \citenamefont {Quan}, \citenamefont {Kim},
  \citenamefont {Fanfarillo}, \citenamefont {Xie}, \citenamefont {Kumar},
  \citenamefont {Park}, \citenamefont {Hemley}, \citenamefont {Vohra},
  \citenamefont {Hennig}, \citenamefont {Hirschfeld}, \citenamefont {Stewart},\
  and\ \citenamefont {Hamlin}}]{Lim_Be22Re_2021}%
  \BibitemOpen
  \bibfield  {author} {\bibinfo {author} {\bibfnamefont {J.}~\bibnamefont
  {Lim}}, \bibinfo {author} {\bibfnamefont {A.~C.}\ \bibnamefont {Hire}},
  \bibinfo {author} {\bibfnamefont {Y.}~\bibnamefont {Quan}}, \bibinfo {author}
  {\bibfnamefont {J.}~\bibnamefont {Kim}}, \bibinfo {author} {\bibfnamefont
  {L.}~\bibnamefont {Fanfarillo}}, \bibinfo {author} {\bibfnamefont {S.~R.}\
  \bibnamefont {Xie}}, \bibinfo {author} {\bibfnamefont {R.~S.}\ \bibnamefont
  {Kumar}}, \bibinfo {author} {\bibfnamefont {C.}~\bibnamefont {Park}},
  \bibinfo {author} {\bibfnamefont {R.~J.}\ \bibnamefont {Hemley}}, \bibinfo
  {author} {\bibfnamefont {Y.~K.}\ \bibnamefont {Vohra}}, \bibinfo {author}
  {\bibfnamefont {R.~G.}\ \bibnamefont {Hennig}}, \bibinfo {author}
  {\bibfnamefont {P.~J.}\ \bibnamefont {Hirschfeld}}, \bibinfo {author}
  {\bibfnamefont {G.~R.}\ \bibnamefont {Stewart}}, \ and\ \bibinfo {author}
  {\bibfnamefont {J.~J.}\ \bibnamefont {Hamlin}},\ }\href {\doibase
  10.1103/PhysRevB.104.064505} {\bibfield  {journal} {\bibinfo  {journal}
  {Phys. Rev. B}\ }\textbf {\bibinfo {volume} {104}},\ \bibinfo {pages}
  {064505} (\bibinfo {year} {2021}{\natexlab{a}})}\BibitemShut {NoStop}%
\bibitem [{\citenamefont {Holzapfel}\ \emph {et~al.}(2001)\citenamefont
  {Holzapfel}, \citenamefont {Hartwig},\ and\ \citenamefont
  {Sievers}}]{Holzapfel_AuEOSatLowT_2001}%
  \BibitemOpen
  \bibfield  {author} {\bibinfo {author} {\bibfnamefont {W.~B.}\ \bibnamefont
  {Holzapfel}}, \bibinfo {author} {\bibfnamefont {M.}~\bibnamefont {Hartwig}},
  \ and\ \bibinfo {author} {\bibfnamefont {W.}~\bibnamefont {Sievers}},\ }\href
  {\doibase 10.1063/1.1370170} {\bibfield  {journal} {\bibinfo  {journal}
  {Journal of Physical and Chemical Reference Data}\ }\textbf {\bibinfo
  {volume} {30}},\ \bibinfo {pages} {515} (\bibinfo {year} {2001})}\BibitemShut
  {NoStop}%
\bibitem [{\citenamefont {Prescher}\ and\ \citenamefont
  {Prakapenka}(2015)}]{Prescher_Dioptas_2015}%
  \BibitemOpen
  \bibfield  {author} {\bibinfo {author} {\bibfnamefont {C.}~\bibnamefont
  {Prescher}}\ and\ \bibinfo {author} {\bibfnamefont {V.~B.}\ \bibnamefont
  {Prakapenka}},\ }\href {https://doi.org/10.1080/08957959.2015.1059835}
  {\bibfield  {journal} {\bibinfo  {journal} {High Pressure Research}\ }\textbf
  {\bibinfo {volume} {35}},\ \bibinfo {pages} {223} (\bibinfo {year}
  {2015})}\BibitemShut {NoStop}%
\bibitem [{\citenamefont {Rietveld}(1969)}]{Rietveld_Rietveld_1969}%
  \BibitemOpen
  \bibfield  {author} {\bibinfo {author} {\bibfnamefont {H.~M.}\ \bibnamefont
  {Rietveld}},\ }\href {\doibase 10.1107/S0021889869006558} {\bibfield
  {journal} {\bibinfo  {journal} {Journal of Applied Crystallography}\ }\textbf
  {\bibinfo {volume} {2}},\ \bibinfo {pages} {65} (\bibinfo {year}
  {1969})}\BibitemShut {NoStop}%
\bibitem [{\citenamefont {{Le Bail}}\ \emph {et~al.}(1988)\citenamefont {{Le
  Bail}}, \citenamefont {Duroy},\ and\ \citenamefont
  {Fourquet}}]{LeBail_LeBaiL_1988}%
  \BibitemOpen
  \bibfield  {author} {\bibinfo {author} {\bibfnamefont {A.}~\bibnamefont {{Le
  Bail}}}, \bibinfo {author} {\bibfnamefont {H.}~\bibnamefont {Duroy}}, \ and\
  \bibinfo {author} {\bibfnamefont {J.}~\bibnamefont {Fourquet}},\ }\href
  {\doibase https://doi.org/10.1016/0025-5408(88)90019-0} {\bibfield  {journal}
  {\bibinfo  {journal} {Materials Research Bulletin}\ }\textbf {\bibinfo
  {volume} {23}},\ \bibinfo {pages} {447 } (\bibinfo {year}
  {1988})}\BibitemShut {NoStop}%
\bibitem [{\citenamefont {Toby}\ and\ \citenamefont
  {Von~Dreele}(2013)}]{Toby_GSASII_2013}%
  \BibitemOpen
  \bibfield  {author} {\bibinfo {author} {\bibfnamefont {B.~H.}\ \bibnamefont
  {Toby}}\ and\ \bibinfo {author} {\bibfnamefont {R.~B.}\ \bibnamefont
  {Von~Dreele}},\ }\href {\doibase 10.1107/S0021889813003531} {\bibfield
  {journal} {\bibinfo  {journal} {Journal of Applied Crystallography}\ }\textbf
  {\bibinfo {volume} {46}},\ \bibinfo {pages} {544} (\bibinfo {year}
  {2013})}\BibitemShut {NoStop}%
\bibitem [{\citenamefont {Jain}\ \emph {et~al.}(2013)\citenamefont {Jain},
  \citenamefont {Ong}, \citenamefont {Hautier}, \citenamefont {Chen},
  \citenamefont {Richards}, \citenamefont {Dacek}, \citenamefont {Cholia},
  \citenamefont {Gunter}, \citenamefont {Skinner}, \citenamefont {Ceder},\ and\
  \citenamefont {Persson}}]{Jain_MaterialsProject_2013}%
  \BibitemOpen
  \bibfield  {author} {\bibinfo {author} {\bibfnamefont {A.}~\bibnamefont
  {Jain}}, \bibinfo {author} {\bibfnamefont {S.~P.}\ \bibnamefont {Ong}},
  \bibinfo {author} {\bibfnamefont {G.}~\bibnamefont {Hautier}}, \bibinfo
  {author} {\bibfnamefont {W.}~\bibnamefont {Chen}}, \bibinfo {author}
  {\bibfnamefont {W.~D.}\ \bibnamefont {Richards}}, \bibinfo {author}
  {\bibfnamefont {S.}~\bibnamefont {Dacek}}, \bibinfo {author} {\bibfnamefont
  {S.}~\bibnamefont {Cholia}}, \bibinfo {author} {\bibfnamefont
  {D.}~\bibnamefont {Gunter}}, \bibinfo {author} {\bibfnamefont
  {D.}~\bibnamefont {Skinner}}, \bibinfo {author} {\bibfnamefont
  {G.}~\bibnamefont {Ceder}}, \ and\ \bibinfo {author} {\bibfnamefont {K.~a.}\
  \bibnamefont {Persson}},\ }\href {\doibase 10.1063/1.4812323} {\bibfield
  {journal} {\bibinfo  {journal} {APL Materials}\ }\textbf {\bibinfo {volume}
  {1}},\ \bibinfo {pages} {011002} (\bibinfo {year} {2013})}\BibitemShut
  {NoStop}%
\bibitem [{\citenamefont {Anzellini}\ \emph {et~al.}(2014)\citenamefont
  {Anzellini}, \citenamefont {Dewaele}, \citenamefont {Occelli}, \citenamefont
  {Loubeyre},\ and\ \citenamefont {Mezouar}}]{Anzellini_ReEOS_2014}%
  \BibitemOpen
  \bibfield  {author} {\bibinfo {author} {\bibfnamefont {S.}~\bibnamefont
  {Anzellini}}, \bibinfo {author} {\bibfnamefont {A.}~\bibnamefont {Dewaele}},
  \bibinfo {author} {\bibfnamefont {F.}~\bibnamefont {Occelli}}, \bibinfo
  {author} {\bibfnamefont {P.}~\bibnamefont {Loubeyre}}, \ and\ \bibinfo
  {author} {\bibfnamefont {M.}~\bibnamefont {Mezouar}},\ }\href {\doibase
  10.1063/1.4863300} {\bibfield  {journal} {\bibinfo  {journal} {Journal of
  Applied Physics}\ }\textbf {\bibinfo {volume} {115}},\ \bibinfo {pages}
  {043511} (\bibinfo {year} {2014})}\BibitemShut {NoStop}%
\bibitem [{\citenamefont {Liu}\ and\ \citenamefont
  {Bassett}(1972)}]{Liu_SrOfcc_1972}%
  \BibitemOpen
  \bibfield  {author} {\bibinfo {author} {\bibfnamefont {L.-G.}\ \bibnamefont
  {Liu}}\ and\ \bibinfo {author} {\bibfnamefont {W.~A.}\ \bibnamefont
  {Bassett}},\ }\href {\doibase https://doi.org/10.1029/JB077i026p04934}
  {\bibfield  {journal} {\bibinfo  {journal} {Journal of Geophysical Research
  (1896-1977)}\ }\textbf {\bibinfo {volume} {77}},\ \bibinfo {pages} {4934}
  (\bibinfo {year} {1972})}\BibitemShut {NoStop}%
\bibitem [{Zen()}]{Zenodo_SrData_2024}%
  \BibitemOpen
  \href {https://zenodo.org/doi/10.5281/zenodo.11093188} {\bibinfo  {journal}
  {All data associated with the present work are available online at
  \url{https://zenodo.org/doi/10.5281/zenodo.11093188}}\ }\BibitemShut
  {NoStop}%
\bibitem [{\citenamefont {Phusittrakool}\ \emph {et~al.}(2008)\citenamefont
  {Phusittrakool}, \citenamefont {Bovornratanaraks}, \citenamefont {Ahuja},\
  and\ \citenamefont {Pinsook}}]{Phusittrakool_SrAbInitio_2008}%
  \BibitemOpen
\bibfield  {journal} {  }\bibfield  {author} {\bibinfo {author} {\bibfnamefont
  {A.}~\bibnamefont {Phusittrakool}}, \bibinfo {author} {\bibfnamefont
  {T.}~\bibnamefont {Bovornratanaraks}}, \bibinfo {author} {\bibfnamefont
  {R.}~\bibnamefont {Ahuja}}, \ and\ \bibinfo {author} {\bibfnamefont
  {U.}~\bibnamefont {Pinsook}},\ }\href {\doibase 10.1103/PhysRevB.77.174118}
  {\bibfield  {journal} {\bibinfo  {journal} {Phys. Rev. B}\ }\textbf {\bibinfo
  {volume} {77}},\ \bibinfo {pages} {174118} (\bibinfo {year}
  {2008})}\BibitemShut {NoStop}%
\bibitem [{\citenamefont {Srepusharawoot}\ \emph {et~al.}(2012)\citenamefont
  {Srepusharawoot}, \citenamefont {Luo}, \citenamefont {Bovornratanaraks},
  \citenamefont {Ahuja},\ and\ \citenamefont
  {Pinsook}}]{Srepusharawoot_Sphase_2012}%
  \BibitemOpen
  \bibfield  {author} {\bibinfo {author} {\bibfnamefont {P.}~\bibnamefont
  {Srepusharawoot}}, \bibinfo {author} {\bibfnamefont {W.}~\bibnamefont {Luo}},
  \bibinfo {author} {\bibfnamefont {T.}~\bibnamefont {Bovornratanaraks}},
  \bibinfo {author} {\bibfnamefont {R.}~\bibnamefont {Ahuja}}, \ and\ \bibinfo
  {author} {\bibfnamefont {U.}~\bibnamefont {Pinsook}},\ }\href {\doibase
  https://doi.org/10.1016/j.ssc.2012.03.035} {\bibfield  {journal} {\bibinfo
  {journal} {Solid State Communications}\ }\textbf {\bibinfo {volume} {152}},\
  \bibinfo {pages} {1172} (\bibinfo {year} {2012})}\BibitemShut {NoStop}%
\bibitem [{\citenamefont {Kim}\ \emph {et~al.}(2012)\citenamefont {Kim},
  \citenamefont {Srepusharawoot}, \citenamefont {Pickard}, \citenamefont
  {Needs}, \citenamefont {Bovornratanaraks}, \citenamefont {Ahuja},\ and\
  \citenamefont {Pinsook}}]{Kim_SrTheory_2012}%
  \BibitemOpen
  \bibfield  {author} {\bibinfo {author} {\bibfnamefont {D.~Y.}\ \bibnamefont
  {Kim}}, \bibinfo {author} {\bibfnamefont {P.}~\bibnamefont {Srepusharawoot}},
  \bibinfo {author} {\bibfnamefont {C.~J.}\ \bibnamefont {Pickard}}, \bibinfo
  {author} {\bibfnamefont {R.~J.}\ \bibnamefont {Needs}}, \bibinfo {author}
  {\bibfnamefont {T.}~\bibnamefont {Bovornratanaraks}}, \bibinfo {author}
  {\bibfnamefont {R.}~\bibnamefont {Ahuja}}, \ and\ \bibinfo {author}
  {\bibfnamefont {U.}~\bibnamefont {Pinsook}},\ }\href {\doibase
  10.1063/1.4742323} {\bibfield  {journal} {\bibinfo  {journal} {Applied
  Physics Letters}\ }\textbf {\bibinfo {volume} {101}},\ \bibinfo {pages}
  {052604} (\bibinfo {year} {2012})}\BibitemShut {NoStop}%
\bibitem [{\citenamefont {Tsuppayakorn-aek}\ \emph {et~al.}(2015)\citenamefont
  {Tsuppayakorn-aek}, \citenamefont {Chaimayo}, \citenamefont {Pinsook},\ and\
  \citenamefont {Bovornratanaraks}}]{Tsuppayakorn-aek_SrIIIMD_2015}%
  \BibitemOpen
  \bibfield  {author} {\bibinfo {author} {\bibfnamefont {P.}~\bibnamefont
  {Tsuppayakorn-aek}}, \bibinfo {author} {\bibfnamefont {W.}~\bibnamefont
  {Chaimayo}}, \bibinfo {author} {\bibfnamefont {U.}~\bibnamefont {Pinsook}}, \
  and\ \bibinfo {author} {\bibfnamefont {T.}~\bibnamefont {Bovornratanaraks}},\
  }\href {\doibase 10.1063/1.4931810} {\bibfield  {journal} {\bibinfo
  {journal} {AIP Advances}\ }\textbf {\bibinfo {volume} {5}},\ \bibinfo {pages}
  {097202} (\bibinfo {year} {2015})}\BibitemShut {NoStop}%
\bibitem [{\citenamefont {Lim}\ \emph {et~al.}(2021{\natexlab{b}})\citenamefont
  {Lim}, \citenamefont {Kim}, \citenamefont {Hire}, \citenamefont {Quan},
  \citenamefont {Hennig}, \citenamefont {Hirschfeld}, \citenamefont {Hamlin},
  \citenamefont {Stewart},\ and\ \citenamefont {Olinger}}]{Lim_A15HP_2021}%
  \BibitemOpen
  \bibfield  {author} {\bibinfo {author} {\bibfnamefont {J.}~\bibnamefont
  {Lim}}, \bibinfo {author} {\bibfnamefont {J.~S.}\ \bibnamefont {Kim}},
  \bibinfo {author} {\bibfnamefont {A.~C.}\ \bibnamefont {Hire}}, \bibinfo
  {author} {\bibfnamefont {Y.}~\bibnamefont {Quan}}, \bibinfo {author}
  {\bibfnamefont {R.~G.}\ \bibnamefont {Hennig}}, \bibinfo {author}
  {\bibfnamefont {P.~J.}\ \bibnamefont {Hirschfeld}}, \bibinfo {author}
  {\bibfnamefont {J.~J.}\ \bibnamefont {Hamlin}}, \bibinfo {author}
  {\bibfnamefont {G.~R.}\ \bibnamefont {Stewart}}, \ and\ \bibinfo {author}
  {\bibfnamefont {B.}~\bibnamefont {Olinger}},\ }\href {\doibase
  10.1088/1361-648X/abeace} {\bibfield  {journal} {\bibinfo  {journal} {Journal
  of Physics: Condensed Matter}\ }\textbf {\bibinfo {volume} {33}},\ \bibinfo
  {pages} {285705} (\bibinfo {year} {2021}{\natexlab{b}})}\BibitemShut
  {NoStop}%
\bibitem [{\citenamefont {Vaks}\ \emph {et~al.}(1988)\citenamefont {Vaks},
  \citenamefont {Samolyuk},\ and\ \citenamefont
  {Trefilov}}]{Vaks_SrAnharmonic_1988}%
  \BibitemOpen
  \bibfield  {author} {\bibinfo {author} {\bibfnamefont {V.}~\bibnamefont
  {Vaks}}, \bibinfo {author} {\bibfnamefont {G.}~\bibnamefont {Samolyuk}}, \
  and\ \bibinfo {author} {\bibfnamefont {A.}~\bibnamefont {Trefilov}},\ }\href
  {\doibase https://doi.org/10.1016/0375-9601(88)90961-9} {\bibfield  {journal}
  {\bibinfo  {journal} {Physics Letters A}\ }\textbf {\bibinfo {volume}
  {127}},\ \bibinfo {pages} {37} (\bibinfo {year} {1988})}\BibitemShut
  {NoStop}%
\bibitem [{\citenamefont {Schilling}(2006)}]{Schilling_AlkaliHP_2006}%
  \BibitemOpen
  \bibfield  {author} {\bibinfo {author} {\bibfnamefont {J.~S.}\ \bibnamefont
  {Schilling}},\ }\href {\doibase 10.1080/08957950600864401} {\bibfield
  {journal} {\bibinfo  {journal} {High Pressure Research}\ }\textbf {\bibinfo
  {volume} {26}},\ \bibinfo {pages} {145} (\bibinfo {year} {2006})}\BibitemShut
  {NoStop}%
\bibitem [{\citenamefont {Lim}\ \emph {et~al.}(2023)\citenamefont {Lim},
  \citenamefont {Sinha}, \citenamefont {Hire}, \citenamefont {Kim},
  \citenamefont {Dee}, \citenamefont {Kumar}, \citenamefont {Popov},
  \citenamefont {Hemley}, \citenamefont {Hennig}, \citenamefont {Hirschfeld},
  \citenamefont {Stewart},\ and\ \citenamefont {Hamlin}}]{Lim_NbMoB2_2023}%
  \BibitemOpen
  \bibfield  {author} {\bibinfo {author} {\bibfnamefont {J.}~\bibnamefont
  {Lim}}, \bibinfo {author} {\bibfnamefont {S.}~\bibnamefont {Sinha}}, \bibinfo
  {author} {\bibfnamefont {A.~C.}\ \bibnamefont {Hire}}, \bibinfo {author}
  {\bibfnamefont {J.~S.}\ \bibnamefont {Kim}}, \bibinfo {author} {\bibfnamefont
  {P.~M.}\ \bibnamefont {Dee}}, \bibinfo {author} {\bibfnamefont {R.~S.}\
  \bibnamefont {Kumar}}, \bibinfo {author} {\bibfnamefont {D.}~\bibnamefont
  {Popov}}, \bibinfo {author} {\bibfnamefont {R.~J.}\ \bibnamefont {Hemley}},
  \bibinfo {author} {\bibfnamefont {R.~G.}\ \bibnamefont {Hennig}}, \bibinfo
  {author} {\bibfnamefont {P.~J.}\ \bibnamefont {Hirschfeld}}, \bibinfo
  {author} {\bibfnamefont {G.~R.}\ \bibnamefont {Stewart}}, \ and\ \bibinfo
  {author} {\bibfnamefont {J.~J.}\ \bibnamefont {Hamlin}},\ }\href {\doibase
  10.1103/PhysRevB.108.094501} {\bibfield  {journal} {\bibinfo  {journal}
  {Phys. Rev. B}\ }\textbf {\bibinfo {volume} {108}},\ \bibinfo {pages}
  {094501} (\bibinfo {year} {2023})}\BibitemShut {NoStop}%
\end{thebibliography}

%

\end{document}